\documentclass[12pt,titlepage]{article}
\usepackage[utf8]{inputenc}
\usepackage{graphicx}
\usepackage{amsthm}
\usepackage{algorithm}
\usepackage{algorithmic}
\usepackage{setspace}

\usepackage{amsmath}

\usepackage[T1]{fontenc}
\usepackage[charter]{mathdesign}

\DeclareMathOperator{\tr}{tr}       
\DeclareMathOperator{\E}{E}               

\title{{\large FEDERAL UNIVERSITY OF MINAS GERAIS \\
Department of Electronic Engineering}
\\[2.5cm]
\Large Technical Report
\\[0.5cm]
\large Robust Bayesian Subspace Identification for Small Data Sets
}

\author{\\[3cm] Alexandre Rodrigues Mesquita}

\begin{document}
\maketitle

\section{Introduction}

Model estimates obtained from traditional subspace identification methods may be subject to significant variance. This elevated variance is aggravated in the cases of large models or of a limited sample size. Common solutions to reduce the effect of variance are regularized estimators, shrinkage estimators and Bayesian estimation. In the current work we investigate the latter two solutions, which have not yet been applied to subspace identification. Our experimental results show that our proposed estimators may reduce the estimation risk up to $40\%$ of that of traditional subspace methods.

In the scope of subspace identification, a number of related works explore regularization techniques such as nuclear norm minimization \cite{verhaegen2016n2sid,smith2014frequency,pillonetto2016regularized,chiuso2019system,sun2022finite}. Under specific hyperparameter choices, regularized estimators can be understood in a Bayesian framework as maximum a posteriori estimators. These regularized estimators carry two important disadvantages in comparison to our approach: i) the choice of hyperparameters is not straightforward and it often does not take advantage of the problem statistics at hand; ii) they are not oriented toward risk minimization.

The usage of regularized, shrinkage and Bayesian estimators in system identification is discussed in \cite{chiuso2016regularization,ljung2020shift,chiuso2019system} and it is greeted as a promising and refreshing trend in this field. We are only aware of one work where a shrinking function is used for subspace identification \cite{liu2018novel}. Differently from our approach, the shrinking is applied to all elements of the matrix and not only to the singular values. In a somewhat related application, in \cite{vajpayee2017data} we find the application of wavelet thresholding as a preprocessor in a subspace identification setting. Apart from the field of system identification, we also find the application of similar methods in the context of subspace identification in hyperspectral imaging \cite{hyperspectral}.

If we regard subspace identification as a particular application of principal component analysis, our work would be understood as an application of robust principal component analysis, which finds many applications in image and video processing and econometrics \cite{candes2011robust}. Bayesian solutions to this problem are found in \cite{elvira2017bayesian,zhao2014robust,babacan2012sparse}. 
Our alternating least squares method was conceived as an improvement upon the Bayesian robust principal component analysis of \cite{ding2011bayesian} and it is closely related to the tensorial regression present in \cite{chu2009probabilistic,gerard2015equivariant,hoff2016equivariant,shi2018alternating}. 

A different approach to subspace identification with finite samples is found in \cite{tsiamis2019finite}, where the focus in on deciding the number of samples in order to bound the estimation error with high probability.

\section{A brief recapitulation of subspace identification}
\label{sec:problem}

Consider the linear state space model in innovation form
\begin{align}
x_{k+1} &= A x_k +B u_k +K e_k \\
y_k &= Cx_k + D u_k + e_k
\enspace,
\end{align}
where $x_k\in \mathbb{R}^{n_x}$ is the state, $y_k\in\mathbb{R}^{n_o}$ is the measured variable, $u_k\in\mathbb{R}^{n_i}$ is the input variable and $e_k$ is the innovations process, assumed to be white Gaussian noise. The model parameters $A, B, C, D$ and $K$ are matrices with the appropriate dimensions.
At times, it may be useful to express the same model in the predictor form
\begin{align}
x_{k+1} &= A_K x_k +B_K z_k \\
y_k &= Cx_k + D u_k + e_k
\enspace,
\end{align}
where $z_k=[u_k^T ~~y_k^T]^T$, $A_K=A-KC$, $B_K=[B-KD~~K]$.

Following the formulation in \cite{qin2006overview}, we make use of the extended state space model
\begin{equation}
Y_f = \Gamma_f X_k + H_f U_f + G_f E_f
\end{equation}
and its predictor form
\begin{equation}
Y_f = H_{fp} Z_p + H_f U_f + G_f E_f
\label{eq:main}
\end{equation}
where the available data is arranged in Hankel matrices defined by
\begin{equation}
Y_f = \left[
\begin{array}{cccc}
y_k & y_{k+1} & \cdots & y_{k+N-1}\\
y_{k+1} & y_{k+2} & \cdots & y_{k+N}\\
\vdots & \vdots & \ddots & \vdots\\
y_{k+f-1} & y_{k+f} & \cdots & y_{k+f+N-2}\\
\end{array}
\right]
\end{equation}
and similarly are defined $U_f$ and $E_f$. Here $f>n$ is the future horizon and $N$ is a function of the size of the available data set. The state sequence is defined as
\begin{equation}
X_k = \left[
\begin{array}{cccc}
x_k & x_{k+1} & \cdots & x_{k+N-1}
\end{array}
\right]
\enspace.
\end{equation}
The past information is collected in $Z_p = [U_p^T ~ Y_p^T]^T$ up to the horizon $p>n$ and arranged as
\begin{equation}
U_p = \left[
\begin{array}{cccc}
u_{k-p} & u_{k-p+1} & \cdots & u_{k-p+N-1}\\
u_{k-p+1} & u_{k-p+2} & \cdots & u_{k-p+N}\\
\vdots & \vdots & \ddots & \vdots\\
u_{k-1} & u_{k} & \cdots & u_{k+N-2}\\
\end{array}
\right]
\label{eq:hankel}
\end{equation}
and likewise for $Y_p$.

As a consequence, $\Gamma_f$ is the extended observability matrix as defined by
\begin{equation}
\Gamma_f =
\left[
\begin{array}{c}
C \\
CA \\
\vdots\\
CA^{f-1}
\end{array}
\right]
\enspace,
\end{equation}
$H_f$ and $G_f$ are Toeplitz matrices given by
\begin{equation}
H_f =
\left[
\begin{array}{cccc}
D & 0 & \cdots & 0 \\
CB & D & \cdots & 0\\
\vdots & \vdots & \ddots & \vdots\\
CA^{f-2}B & CA^{f-3}B & \cdots & D
\end{array}
\right]
\enspace,
\end{equation}
and
\begin{equation}
G_f =
\left[
\begin{array}{cccc}
I & 0 & \cdots & 0 \\
CK & I & \cdots & 0\\
\vdots & \vdots & \ddots & \vdots\\
CA^{f-2}K & CA^{f-3}K & \cdots & I
\end{array}
\right]
\cdot I_f\otimes \Sigma^{1/2}
\enspace,
\label{eq:lowtoep}
\end{equation}
where $\Sigma$ is the innovations covariance.
The matrix $H_{fp}$ is reminiscent of the matrix of Markov parameters and is given by 
\begin{equation}
H_{fp} = \left[
\begin{array}{cc}
H_{fp}^{(1)} & H_{fp}^{(2)}
\end{array}
\right]
\end{equation}
with 
\begin{equation}
H_{fp}^{(i)} = 
\left[
\begin{array}{cccc}
CA_K^{p-1}B_K^{(i)} & CA_K^{p-2}B_K^{(i)} & \cdots & CB_K^{(i)} \\
CAA_K^{p-1}B_K^{(i)} & CAA_K^{p-2}B_K^{(i)} & \cdots & CB_K^{(i)} \\
\vdots & \vdots & \ddots & \vdots \\
CA^{f-1}A_K^{p-1}B_K^{(i)} & CA^{f-1}A_K^{p-2}B_K^{(i)} & \cdots & CA^{f-1}B_K^{(i)} \\
\end{array}
\right]
\enspace,
\end{equation}
for $i=1,2$ and $B_K^{(1)}=B-KD$ and $B_K^{(2)}=K$. These matrices can also be decomposed as products $H_{fp}^{(i)} = \Gamma_fL_p^{(i)}$ of the extended observability $\Gamma_f$ and controllability matrices:
\begin{equation}
L_p^{(i)}=[
\begin{array}{ccc}
A_K^{p-1}B_K^{(i)} & A_K^{p-2}B_K^{(i)} \cdots B_K^{(i)}
\end{array}
]
\enspace
\end{equation}
for $i=1,2$.

Several estimation methods were proposed in the literature departing from this formulation and exploiting the fact that the data as structured in (\ref{eq:main}) should lie in low dimensional subspaces (see \cite{qin2006overview,van2012subspace} for a comprehensive review). A common approach is to solve (\ref{eq:main}) via least squares and then estimate $\Gamma_f$ from a truncated singular value decomposition of $H_{fp}$. The matrices $A$ and $C$ could then be recovered from estimates of $\Gamma_f$ and $\Gamma_{f+1}$. The matrices $B$ and $D$ could then be obtained from a new least squares problem or from estimates of $L_p$.

From a general perspective, most methods obtain an initial estimate $\hat{H}_{fp}$ from least squares and, given two weight matrices $W_1$ and $W_2$, obtain a low-dimensional estimate $\hat{\hat{H}}_{fp}$ by truncating the singular value decomposition of ${\hat{H}}_{fp}$ to the largest $r$ components:
\begin{equation}
W_1\hat{H}_{fp}W_2 = USV^T \approx U_r S_r V_r^T =: W_1\hat{\hat{H}}_{fp}W_2
\enspace.
\end{equation}
In this context, $r$ would be an estimate of the system order $n$. Methods of obtaining such an estimate of $n$ have been treated with difficulty in the literature and usually one resorts to ad hoc solutions or the visual inspection of the singular values.

Traditional subspace identification methods obtain an initial estimate of $H_{fp}$ from (\ref{eq:main}) applying the least squares method as follows:
\begin{equation}
[\hat{H}_{fp} ~\hat{H}_f] = Y_f\left[
\begin{array}{c}
U_p\\
Y_p\\
U_f
\end{array}
\right]^\dagger
\enspace.
\label{eq:LS}
\end{equation}
Notice, however, that this estimate only approximates the maximum likelihood estimate as the noise term $G_fE_f$ in (\ref{eq:main}) is at best only approximately white and as $H_f$ is a lower triangular Toeplitz matrix.

At this point it is worthwhile having a discussion on the approximations typically made in the process of solving (\ref{eq:main}). Given that $E_f$ is structured as a Hankel matrix as in (\ref{eq:hankel}), its components are not truly independent. Nevertheless, the off diagonal elements of $\text{E}[E_fE_f^T]$ are reasonably sparse, which gives a reasonable approximation of a white matrix. In order the incorporate the Hankel and Toeplitz structure of $E_f$ and $H_f$ in the computation, one would have to vectorize equation (\ref{eq:main}):
\begin{equation}
y_f = (Z_p^T \otimes I) h_{fp} + (U_f^T\otimes I) h_f + (I\otimes G_f) e_f
\enspace,
\end{equation} 
where the lower case letters denote the vectorized version of the corresponding matrices. In addition, $h_f$ and $e_f$ lie in lower dimensional subspaces. This can be solved introducing known matrices $B_{T}$ and $B_{W}$ that convert the original lower dimensional data into the vectorized versions of the Toeplitz matrix $H_f$ and the Hankel matrix $E_f$. This would result in the new equation
\begin{equation}
y_f = (Z_p^T \otimes I) h_{fp} + (U_f^T\otimes I) B_{T} \bar{h}_f + (I\otimes G_f) B_{W}\bar{e}_f
\enspace.
\label{eq:vectorized}
\end{equation} 
However, the least squares problem that arises in (\ref{eq:vectorized}) is high dimensional, sparse and poorly conditioned numerically. In this regard, one of the virtues of the formulation in (\ref{eq:main}) is to avoid the numerical difficulties that would arise in a more detailed formulation.

\section{Problem Description}

Most subspace identification methods rely on the limit case of large enough data sets, $N\to\infty$. In addition, most approaches are based on least squares or, equivalently, on maximum likelihood estimators of $H_{fp}$. However, it is well known that such estimators perform poorly in regards to the estimation risk under quadratic loss (are not admissible) when the number of parameters is larger than 2 \cite{lehmann2006theory,robert2007bayesian}, which is classically known as the Stein phenomenon \cite{james1961estimation}. On the other hand, the number of parameters in $H_{fp}$ may be as large as $n_o\cdot(n_i+n_o)\cdot f\cdot p$, which may be considerably large. Even if we consider that $H_{fp}$ is low rank and may be parameterized and estimated in a smaller space, it would still depend on $(n_i+2n_o)\cdot n_x$ parameters, which is still large in most applications.

In addition, there seems to be considerable difficulty in the literature to find appropriate methods to estimate the system order $n_x$, which is of particular concern given the high sensitivity of the results to this parameter.   

Within this context, our goal is to investigate robustified methods to estimate $H_{fp}$. In particular, two groups of methods shall be considered. The first group is comprised of the main shrinkage estimators of singular values available in the literature. The second group consists of Bayesian estimators that can be efficiently computed with a Gibbs sampling scheme.

Given that the estimation errors in the matrices $(A,B,C,D,K)$ are upper bounded by a constant times $\|H_{fp}-\hat{H}_{fp}\|$ as demonstrated in \cite{tsiamis2019finite}, we concentrate our efforts on the estimators of $H_{fp}$ and leave an investigation on the matrices estimators to future work.

\section{Singular Value Shrinkage}

Shrinkage estimators are biased estimators that shrink the maximum likelihood estimate towards a pre-assigned value (often to zero). These estimators were first proposed by Stein \cite{james1961estimation} and it has been demonstrated that this type of estimator may provide smaller estimation risk than the maximum likelihood estimator under quadratic loss when the number of parameters is larger than 2. The rationale behind such estimators is that, although shrinking introduces bias, it reduces variance and therefore it may be beneficial in reducing the overall quadratic error.

In the context of singular value decompositions, approaches based on random matrix theory obtain optimal shrinkage estimators in the asymptotic case of the matrix dimensions approaching infinity. This assumption may be far from adequate in the case of subspace identification with small data sets. Non-asymptotic results are also available based on the minimization of unbiased estimates. 

Most approaches consider the problem of estimating some matrix $X\in\mathbb{R}^{i\times j}$ from a noisy measurement
\begin{equation}
Y = X + \sigma	W
\enspace,
\label{eq:svdmodel}
\end{equation}
where $\sigma>0$ is known and the noise matrix $W$ is orthogonally invariant, i.e., its distribution is the same as that of $UWV$ for any orthogonal matrices $U$ and $V$. In practice, $Y$ is not a single measurement, but the maximum likelihood estimate obtained from multiple measurements, which gives a sufficient statistic based on all available measurements.

Considering the Frobenius norm squared as the risk loss function, and given that the noise is orthogonally invariant, it suffices to restrict our search to shrinkage estimators of the form:
\begin{equation}
\hat{X} = U \eta(S) V^T
\enspace,
\end{equation}
where $U S V^T$ is the singular value decomposition of $Y$ and $\eta(\cdot)$ is the shrinkage function.

\subsection{Hard and Soft-Thresholding}

In the asymptotic case of very large dimensions, \cite{gavish2014optimal} computed the optimal parameters for the hard and soft-thresholding functions:
\begin{align}
\eta_{\text{hard}} (s) &= s 1_{s>\lambda_{\text{hard}}}  \\
\eta_{\text{soft}} (s) &= (s-\lambda_{\text{soft}})_+  
\enspace,
\end{align} 
where $1_{(\cdot)}$ is the indicator function and $(\cdot)_+$ is the positive-part function and where the function $\eta$ are applied element-wise on the singular values and the optimal thresholds are
\begin{equation}
\lambda_{\text{hard}} =  \sqrt{2(\beta+1)+\frac{8\beta}{\beta+1+\sqrt{\beta^2+14\beta+1}}}\cdot\sigma\sqrt{j}
\end{equation}
and
\begin{equation}
\lambda_{\text{soft}} = (1+\sqrt{\beta})\sigma\sqrt{j}
\enspace,
\end{equation}
with $\beta=i/j$ being the aspect ratio of the matrix $Y$.

\subsection{Optimal shrinkage in the asymptotic case}

Still in the asymptotic case, the overall optimal shrinkage function was obtained in \cite{gavish2017optimal}. For the squared Frobenius norm loss function, the optimal shrinkage is given by
\begin{equation}
\eta^*(s) = \frac{1}{s}\sqrt{(s^2-(1+\beta)\sigma^2j)^2-4\beta\sigma^4j^2} \cdot 1_{s>\lambda_{\text{soft}}}
\enspace.
\end{equation}

\subsection{SURE-based thresholding}

In the non-asymptotic case, \cite{candes2013unbiased} obtained an expression for the Stein's unbiased risk estimate (SURE) for estimators based on soft-thresholding functions. The threshold values can then be obtained by minimizing the SURE risk estimate. For $\eta(s) = (s-\lambda)_+$ and the Frobenius norm squared loss, the unbiased risk estimate is given by 
\begin{multline}
\text{SURE}_\lambda(Y) = -ij\sigma^2 + \sum_{k=1}^{i} \min(\lambda^2,\sigma_k^2) +
2\sigma^2\left[
(j-i)\sum_{k=1}^{i} \left(1-\frac{\lambda}{\sigma_k}\right)_+
+ \sum_{k=1}^{i} 1_{\sigma_k>\lambda} \right.\\
\left.
+ \sum_{k,l,k\neq l}^i\frac{\sigma_k(\sigma_k-\lambda)_+}{\sigma_k^2-\sigma_l^2}
\right]
\enspace,
\end{multline}
where $\sigma_k$ denotes the $k$-th singular value of $Y$ and $i<j$. Given $Y$, $\text{SURE}_\lambda(Y)$ can be easily minimized since it is a piecewise quadratic function in $\lambda$.

\subsection{Subspace identification with shrinkage estimators}

Following the traditional subspace identification methods, we obtain the initial estimate of $H_{fp}$ using the least squares estimate of (\ref{eq:LS}). This estimate offers the advantage of being unbiased and to have a covariance that can be easily characterized. This facilitates framing the problem of singular value shrinking in the format given by (\ref{eq:svdmodel}).

In the context of shrinkage estimators, one could naturally propose to also perform shrinkage of the parameters estimated in (\ref{eq:LS}) by replacing the linear regression by a ridge regression. However, we would have a biased estimate of $H_{fp}$ and this would make it harder to explore the theory of shrinkage estimators for singular values.

Our goal is to minimize the risk function
\begin{equation}
\mathcal{R}(\hat{\hat{H}}_{fp}) = \text{E}\left[\tr\left(W_2^T(H_{fp}-\hat{\hat{H}})^TW_1^TW_1(H_{fp}-\hat{\hat{H}})W_2\right)\right]
\enspace.
\label{eq:risk}
\end{equation}

Within the domain of the shrinkage estimators presented above, we shall minimize the risk in (\ref{eq:risk}) if we shrink the SVD decomposition of $W_1\hat{H}_{fp}W_2$. However, the variance level $\sigma$ is assumed known in (\ref{eq:svdmodel}). This means that we still must obtain a reasonable estimate for it.

Defining the projection matrix onto the orthogonal complement of the row space of $U_f$ as $\Pi_{U_f}^\perp = I - U_f^T(U_fU_f^T)^{-1}U_f$, the estimate $\hat{H}_{fp}$ in (\ref{eq:LS}) is equivalent to \cite{qin2006overview}
\begin{equation}
\hat{H}_{fp} = Y_f \Pi_{U_f}^\perp	Z_p^T(Z_p\Pi_{U_f}^\perp Z_p^T)^{-1}
\enspace.
\end{equation}
Substituting $Y_f$ as in (\ref{eq:main}), this leads to
\begin{equation}
\hat{H}_{fp} = H_{fp}+G_fE_f \Pi_{U_f}^\perp	Z_p^T(Z_p\Pi_{U_f}^\perp Z_p^T)^{-1}
\enspace.
\end{equation}
Multiplying by the weight matrices, we have
\begin{equation}
W_1\hat{H}_{fp} W_2 = W_1H_{fp}W_2+W_1 G_fE_f \Pi_{U_f}^\perp	Z_p^T(Z_p\Pi_{U_f}^\perp Z_p^T)^{-1}W_2
\enspace.
\label{eq:noise}
\end{equation}

Unless we carefully pick the weight matrices, the noise term in (\ref{eq:noise}) is not orthogonally invariant white as assumed in (\ref{eq:svdmodel}). If we consider the approximation $\E[e_fe_f^T]=I$ for the vectorized noise $E_f$, then the covariance matrix for the vectorization of $W_1\hat{H}_{fp}W_2$ would be
\begin{equation}
\Sigma_{h_{fp}} = W_2^T(Z_p\Pi_{U_f}^\perp Z_p^T)^{-1}W_2\otimes W_1 G_fG_f^TW_1^T
\enspace.
\end{equation}
To make this covariance orthogonally invariant, one would choose $W_1=(G_fG_f^T)^{-1/2}$ and $W_2=(Z_p\Pi_{U_f}^\perp Z_p^T)^{1/2}$. This is closely related to the weight choice of the popular CVA algorithm for subspace identification \cite{qin2006overview}. Nonetheless, there might be reasons to choose different weights. For example, the N4SID algorithm uses $W_1=I$ and $W_2=Z_p$ in order to penalize the prediction error. Alternatively, one might prefer to penalize errors in the in the input-output relation and choose $W_1=I$ and $W_2=I$.

For general weight matrices, we identify the noise level $\sigma$ with the largest possible noise level in a given direction, i.e.,
\begin{equation}
\sigma^2 = \sigma_{\max}(W_2^T(Z_p\Pi_{U_f}^\perp Z_p^T)^{-1}W_2) \cdot \sigma_{\max}(W_1 G_fG_f^TW_1^T)
\enspace.
\label{eq:sigma}
\end{equation}

In order to estimate $G_f$, we compute the residues 
\begin{equation}
\mathcal{E} = Y_f - \hat{H}_{fp}Z_p - \hat{H}_fU_f
\label{eq:residues}
\end{equation}
and construct the estimate
\begin{equation}
\hat{G}_f\hat{G}_f^T = \frac{\mathcal{E}\mathcal{E}^T}{j-i(n_o+2n_i)} 
\label{eq:estG}
\end{equation}
where we make $f=i$ and $N=j$ and $i(n_o+2n_i)$ is the number o degrees of freedom of the linear regression. The estimate in (\ref{eq:estG}) is enough to obtain $\sigma$ from (\ref{eq:sigma}). If we further desire to specify $G_f$, one simple approach is to obtain the Cholesky decomposition of (\ref{eq:estG}) and then average over the matrix diagonals to enforce the Toeplitz structure on $\hat{G}_f$. If an initial truncated estimate $\hat{\hat{H}}_{fp}$ is computed, it may be used in (\ref{eq:residues}) to further account for the noise that is present in the lower singular values of $\hat{H}_{fp}$. In that case, the number of degrees of freedom could be changed to $in_i+i(n_i+n_o)-(i(n_i+n_o)-(i+n_i+n_o)r+r^2)$, where $r$ is the rank of $\hat{\hat{H}}_{fp}$.

\section{An Alternating Least Squares Bayesian Approach}

Bayesian estimators are inherently robust. Under mild conditions \cite{robert2007bayesian}, they are admissible, i.e., there exists no estimator that improves their risk for all the parameters in the parameter space. Under group invariance of priors and losses (see \cite{robert2007bayesian} for definitions), they are also minimax, i.e., they minimize the risk under the worst case parameter value. The shrinkage estimators presented above may often be characterized as Bayes estimators with an empirical prior, i.e., a prior distribution that is constructed from the data itself.

In our approach to subspace identification, we aim to construct a Bayesian method that is computationally simple. With this in mind, we choose priors that lead to simple regularized least squares steps. The priors are obtained empirically from the data. We have experimented with hierarchical Bayes priors as well but did not find a convincing improvement with respect to the simpler empirical priors. Our model was inspired by that in \cite{ding2011bayesian} but, differently from this paper, we do not explicitly estimate the orthogonal matrices $U$ and $V$ and the singular value matrix $S$. We found that better results are obtained by estimating $US^{1/2}$ and $S^{1/2}V^T$ instead.

For the noise term $G_f$, we propose a prior invariant with respect to the group of lower triangular Toeplitz matrices. Hence, the covariance estimator will be equivariant with respect to this group. In this particular case, it is possible to compute posterior samples in a reasonably simple manner.

Assuming $n_i=n_o=1$ for simplicity, we adopt the following set of independent priors:

\begin{align}
\Gamma_f &= \bar{G}_f \Xi_\Gamma \Lambda_\Gamma^{-1/2} \label{eq:gammaprior}\\
H_f &= \bar{G}_f \Xi_H \Lambda_H^{-1/2} \\
L_p &= \Lambda_L^{-1/2} \Xi_L Z_p^\dagger \label{eq:lpprior}\\
G_f &\sim \frac{1}{|G_{f[1,1]}|^{i}}
\\
\end{align}
where the $\Xi_{(\cdot)}$ matrices are random matrices with the appropriate dimensions and whose elements are independent normally distributed random variables with variance $1$, the $\Lambda_{(\cdot)}$ matrices are fixed parameters to be specified later and $\bar{G}_f=G_f/G_{f[1,1]}$. 
It will be shown later that the improper prior given to $G_f$ is invariant under the group of multiplication by lower triangular Toeplitz matrices. 

Since the full posterior distribution for the estimation problem at hand is too hard to characterize analytically, we estimate its empirical distribution using a Gibbs sampler. In a Gibbs sampler, samples from dependent variables $x$ and $y$ are drawn iteratively from their conditional distributions as in: \begin{align}
&y^{(n)}\sim p(y|x^{(n-1)}) \\
&x^{(n)}\sim p(x|y^{(n)})
\enspace.
\end{align}
The probability distribution of the resulting Markov chain $(x^{(n)},y^{(n)})$ is shown to converge under mild conditions to $p(x,y)$. The following posterior updates follow from our choice of prior:
\begin{multline}
[
\begin{array}{cc}
\Gamma_f^{(n)} & H_f^{(n)}
\end{array}]
 = Y_f
 \left[
 \begin{array}{c}
 X_p^{(n-1)} \\
 U_f
\end{array}  
\right]^T
\left(
 \left[
 \begin{array}{c c}
 \Lambda_\Gamma & 0\\
 0 & \Lambda_H \\
\end{array}  
\right]+
\gamma^{(n)}
 \left[
 \begin{array}{c}
 X_p^{(n-1)} \\
 U_f
\end{array}  
\right]
 \left[
 \begin{array}{c}
 X_p^{(n-1)} \\
 U_f
\end{array}  
\right]^T
\right)^{-1}
\gamma^{(n)}
\\+
\bar{G}_f^{(n-1)}\Xi_{\Gamma,H}^{(n)}\left(
 \left[
 \begin{array}{c c}
 \Lambda_\Gamma & 0\\
 0 & \Lambda_H \\
\end{array}  
\right]+
\gamma^{(n)} \left[
 \begin{array}{c}
 X_p^{(n-1)} \\
 U_f
\end{array}  
\right]
 \left[
 \begin{array}{c}
 X_p^{(n-1)} \\
 U_f
\end{array}  
\right]^T
\right)^{-1/2}
\label{eq:reg1}
\end{multline}
and
\begin{multline}
L_p^{(n)}
 = 
\left(
\Gamma_f^{(n)T}(\Sigma_e^{(n)})^{-1}\Gamma_f^{(n)}
+\Lambda_L\right)^{-1}
\Gamma_f^{(n)T}(\Sigma_e^{(n)})^{-1}
(Y_f-H_f^{(n)}U_f)\\
+
\left(
\Gamma_f^{(n)T}(\Sigma_e^{(n)})^{-1}\Gamma_f^{(n)}
+\Lambda_L\right)^{-1/2}\Xi_L^{(n)}Z_p^\dagger
\label{eq:reg2}
\end{multline}
where $X_p^{(n)}=L_p^{(n)}Z_p$, $\gamma^{(n)}=1/(G_{f[1,1]}^{(n-1)}])^2$ and $\Sigma_e^{(n)} = G_f^{(n-1)}(G_f^{(n-1)})^T$, the $\Xi_{(\cdot)}^{(n)}$ matrices are random matrices whose components are independently drawn from a unit normal distribution. Intuitively, we are performing independent regressions row by row in (\ref{eq:reg1}) and independent regressions column by column in (\ref{eq:reg2}), in addition to summing the corresponding simulated noise terms. 

In order to define the posterior update for $G_f$, we first define the matrix $B_T$ that maps the last row of $G_f$ to $\text{vec}(G_f)$, i.e., $\text{vec}(G_f) = B_T (G_{f[i,\cdot]})^T$. Similarly, we define the matrix $B_W$ such that $\text{vec}(E_f)=B_W e_{[1:i+j-1]}$. Let $\chi_k$ denote the chi distribution with $k$ degrees of freedom, then the posterior update for $G_f$ is given by:
\begin{align}
\mathcal{E}^{(n)} &= Y_f - \Gamma_f^{(n)}L_p^{(n)}Z_p -H_f^{(n)}U_f \label{eq:posti}\\
\Omega^{(n)} &= B_T^T(\mathcal{E}^{(n)}\otimes I_i)B_W(B_W^TB_W)^{-1}B_W^T((\mathcal{E}^{(n)})^T\otimes I_i)B_T \label{eq:Omega}\\
\nu_i^{(n)} &\sim \chi_{j+1} \\
\nu_k^{(n)} &\sim N(0,1) , ~\text{for} ~k=1, \ldots, i-1\\
(G_f^{(n)})^{-1}_{[i,:]} &= \nu_{[1:i]}(\Omega_L^{(n)})^{-1} \\
\text{vec}\left((G_f^{(n)})^{-1}\right) &= B_T(G_f^{(n)})^{-1}_{[i,:]} \label{eq:postf}
\enspace,
\end{align}
where $\Omega_L^{(n)}$ denotes the lower triangular part of the Cholesky decomposition of $\Omega^{(n)}$.
In other words, we first obtain a sample of the last row of $G_f^{-1}$, then we construct the full lower triangular Toeplitz matrix $G_f^{-1}$ from this row and compute its inverse. Given that the matrix $B_W$ is very large and sparse, (\ref{eq:Omega}) may be somewhat tricky to compute. A safer alternative may come from ignoring the Hankel structure of $E_f$ and assuming that its elements are mutually independent. In this case, the posterior update becomes
\begin{align}
\mathcal{E}^{(n)} &= Y_f - \Gamma_f^{(n)}L_p^{(n)}Z_p -H_f^{(n)}U_f \label{eq:posti2}\\
\Omega^{(n)} &= B_T^T(\mathcal{E}^{(n)}(\mathcal{E}^{(n)})^T\otimes I_i)B_T \\
\nu_i^{(n)} &\sim \chi_{ij-i+2} \\
\nu_k^{(n)} &\sim N(0,1) , ~\text{for} ~k=1, \ldots, i-1\\
\left[(G_f^{(n)})^{-1}\right]_{i,\cdot} &= \nu_{[1:i]}(\Omega_L^{(n)})^{-1} \\
\text{vec}\left((G_f^{(n)})^{-1}\right) &= B_T(G_f^{(n)})^{-1}_{[i,\cdot]} \label{eq:postf2}
\enspace.
\end{align}
Finally, the estimate for ${H}_{fp}$ is obtained by averaging the over the trajectory of the Markov chain:
\begin{equation}
\hat{H}_{fp} = \frac{1}{N_F-N_o} \sum_{n=N_o}^{N_F} \Gamma_f^{(n)}L_p^{(n)}
\enspace,
\end{equation} 
where $N_o$ is some burn-in period intended to remove the effect of transients. In order to reduce variance and improve convergence, one may prefer to average over the expected values of (\ref{eq:reg1}) and (\ref{eq:reg2}) in every step (obtained by setting the respective $\Xi_{(\cdot)}$ matrices to zero):
\begin{equation}
\hat{H}_{fp} = \frac{1}{2(N_F-N_o)} \sum_{n=N_o}^{N_F} \E[\Gamma_f^{(n)}]L_p^{(n-1)}+\Gamma_f^{(n)}\E[L_p^{(n)}]
\enspace.
\end{equation}
Regarding the parameters in the priors, we initially obtain estimates ${H}_{fp}^{(1)}$ and ${H}_f^{(1)}$ from (\ref{eq:LS}) and next some truncated singular value decomposition $H_{fp}^{(1)}Z_p \approx U_rS_rV_r^T$. Then, we make $G_f^{(1)}=I_i$ and
\begin{align}
\Gamma_f^{(1)} &= U_rS_r^{1/2}\\
L_p^{(1)} &=S_r^{1/2}V_r^TZ_p^\dagger \\
\Lambda_\Gamma^{-1} &=S_r/i \label{eq:Lambda}\\
\Lambda_L^{-1} &=S_r/j \\
\Lambda_H^{-1} &= I_{i}\tr( (H_f^{(1)})^TH_f^{(1)})/i^2
\enspace.
\end{align}
To justify such a choice, we note that the priors (\ref{eq:gammaprior}) and (\ref{eq:lpprior}) approximately describe an SVD decomposition:
\begin{equation}
\Gamma_f X_p = \frac{\Xi_\Gamma}{i} S_r\frac{\Xi_L}{j}
\end{equation}
where ${\Xi_\Gamma}/{i}$ and ${\Xi_L}/{j}$ behave as orthogonal matrices in expectation, i.e., $\E[\Xi_\Gamma^T\Xi_\Gamma/{i^2}]= \E[\Xi_L\Xi_L^T/{j^2}] = I_r$.

\section{Numerical Experiments}

In order to test our proposed methods, we ran Monte Carlo simulations on a large number of systems and compared the estimation risks.

The system order $n_x$ was uniformly distributed from $1$ to $10$ and we fixed $n_i=n_o=1$. The sample size was defined as $N=\lfloor 80\sqrt{n_x}\rfloor$. The row-length of Hankel matrices was set as $i=\lfloor N/10\rfloor$. 

To obtain a stable system matrix $A$, we first sampled an auxiliary matrix $\tilde{A}$ such that $\tilde{A}_{[k,l]}\sim N(0,1)$ and a spectral radius $\lambda_a\sim \mathcal{U}(0,1)$. Then we made $A=\tilde{A}/\lambda_{\max}(\tilde{A})\cdot \lambda_a$. The matrices $B$ and $C$ were sampled such that $B_{[k,l]},C_{[k,l]}\sim N(0,1)$. We make $D=0$. The measurement noise covariance $R_v$ was generated such that $(R_v)^{1/2}_{[k,l]}\sim N(0,1)$. Likewise, the process noise covariance $R_w$ was generated such that $(R_w)^{1/2}_{[k,l]}\sim N(0,1)$. The Kalman gain $K$ was therefore obtained from the previous parameters.

The system input $u_k$ is comprised of independent samples from $N(0,\textsf{SNR})$, where the signal to noise ratio \textsf{SNR} was sampled uniformly on logarithmic scale such that $\log_{10}\textsf{SNR}\sim \mathcal{U}(-1,2)$.

We applied the proposed algorithms to each model realization and response realization. For each realization and each estimator, we computed the risk
\begin{equation}
\mathcal{R}_k = \tr\left(W_2^T(H_{fp}-\hat{\hat{H}}_{fp})^TW_1^TW_1(H_{fp}-\hat{\hat{H}}_{fp})W_2\right)
\enspace.
\end{equation}

Given that the system models have different scales in each realization, we normalized the risk performance $\mathcal{R}_k$ by that of a reference estimator $\mathcal{R}_k^o$ and averaged over \textsf{N} realizations as such:
\begin{equation}
\bar{\mathcal{R}}=\exp\left(\frac{1}{\textsf{N}}\sum_{k=1}^\textsf{N}\ln\left(\frac{\mathcal{R}_k}{{\mathcal{R}}_k^o}\right)\right)
\enspace.
\end{equation}
The logarithmic function is used to give higher and symmetric weights to risks that are either too low or too high compared to the reference.

We used the same estimate of $\hat{G}_f$ for all shrinkage methods as defined by (\ref{eq:estG}). We started by truncating $\hat{H}_{fp}$ to the largest $r$ singular values and then, computing the corresponding $\hat{G}_f(r)$ and hard threshold $\lambda_\text{soft}(r)$, we obtained
\begin{equation}
r^* = \min\{r:\max\{l:S_{[l]}>\lambda_\text{soft}(r)\}<r\}
\enspace.
\end{equation}
In words, $r^*$ is the least $r$ such that the order estimate is less than $r$. We then make $\hat{G}_f=\hat{G}_f(r^*)$ and use $r^*$ in (\ref{eq:estG}).

\subsection{A heuristic as benchmark}

To provide the reference estimator above, we propose a heuristic that seeks to mimic the order selection as done by visual inspection. Namely, what a typical user would do is to look at the plot of singular values and identify the point of sharp change in their rate of decline. In order to do something similar automatically, we borrow from the idea of effective sample size and, from the vector $S$ of ordered singular values, define
\begin{equation}
n_{eff} = \frac{\left(\sum_{l=1}^i S_l\right)^2}{\sum_{l=1}^i S_l^2}
\enspace.
\label{eq:heuristic1}
\end{equation}
Next we construct a function $\eta(l)$ that linearly fits $\ln S_l$ from $l=n_{eff}+1$ to $l=i$. Finally we define our heuristic order estimate as
\begin{equation}
\hat{n}_x = \max \{l: S_l>\exp(\eta(l))\} 
\enspace.
\label{eq:heuristic2}
\end{equation}

As a second benchmark, we also considered the selection criterion adopted in \cite{verhaegen2016n2sid,ljung2007system}:
\begin{equation}
\hat{n}_{x,2} = \max \{l: S_l>\exp((\ln S_1 +\ln S_i)/2)\} 
\enspace.
\end{equation}

\subsection{Results}

Our results are summarized in Tables \ref{tab:tab1}, \ref{tab:tab2} and \ref{tab:tab3}, where we considered the three main weight choices. We observe that the optimal shrinkage method and the alternating least squares approach give consistently lower risk estimates. Interestingly, we observe that soft-thresholding and hard-thresholding do not always improve upon the benchmark. Since the SURE-based method also applies soft-thresholding, we see that the problem does not lie in the class of shrinking functions, but on a poor parameter choice that was based on asymptotic properties of large random matrices. 

\begin{table}
\centering
\begin{tabular}{|c|c|}
\hline 
Method & Average Normalized Risk ($\pm 5\%$) \\ 
\hline 
Heuristic (\ref{eq:heuristic1}) & 1.0 \\ 
\hline 
Heuristic (\ref{eq:heuristic2}) & 3.83\\
\hline
Hard-thresholding & 0.97\\
\hline
Soft-thresholding & 0.76\\
\hline
Optimal Shrinkage & 0.75\\
\hline
SURE & 0.58 \\
\hline
Alternating Least Squares & 0.39\\
\hline
\end{tabular} 
\caption{Risk performance for $W_1=I_i$ and $W_2=I_{2i}$ and $3000$ Monte Carlo runs.  Other paremeters are $N_F=250$, $N_o=1$.}
\label{tab:tab1}
\end{table}

\begin{table}
\centering
\begin{tabular}{|c|c|}
\hline 
Method & Average Normalized Risk ($\pm 5\%$) \\ 
\hline 
Heuristic (\ref{eq:heuristic1}) & 1.0 \\ 
\hline 
Heuristic (\ref{eq:heuristic2}) & 1.53\\
\hline
Hard-thresholding & 0.75\\
\hline
Soft-thresholding & 1.57\\
\hline
Optimal Shrinkage & 0.58\\
\hline
SURE & 0.73 \\
\hline
Alternating Least Squares & 0.52\\
\hline
\end{tabular} 
\caption{Risk performance for $W_1=\hat{G}_f^{-1}$ and $W_2=(Z_p\Pi_{U_f}^\perp Z_p^T)^{1/2}$ (similarly to CVA) and $3000$ Monte Carlo runs. Other parameters are $N_F=250$, $N_o=1$.}
\label{tab:tab2}
\end{table}

\begin{table}
\centering
\begin{tabular}{|c|c|}
\hline 
Method & Average Normalized Risk ($\pm 5\%$) \\ 
\hline 
Heuristic (\ref{eq:heuristic1}) & 1.0 \\ 
\hline 
Heuristic (\ref{eq:heuristic2}) & 1.52\\
\hline
Hard-thresholding & 0.68\\
\hline
Soft-thresholding & 1.20\\
\hline
Optimal Shrinkage & 0.49\\
\hline
SURE & 0.64 \\
\hline
Alternating Least Squares & 0.42\\
\hline
\end{tabular} 
\caption{Risk performance for $W_1=I_i$ and $W_2=Z_p$ (as in N4SID) and $3000$ Monte Carlo runs. Other parameters are $N_F=250$, $N_o=1$.}
\label{tab:tab3}
\end{table}

\appendix

\section{Equivariant Estimators of the Covariance on the Lower Triangular Toeplitz Group}

The set of lower triangular Toeplitz matrices as exemplified in (\ref{eq:lowtoep}) is a group under matrix multiplication. This group operation may be interpreted as the cascading of dynamical systems. In this sense, we are interested in estimators that are invariant with respect to dynamical system cascading. For example, if we pass both input and output through a linear filter, we want an estimator that gives the same model regardless of the filtering. In the Bayesian framework, not all priors result in such an equivariant estimator. In this section, we aim at deriving a prior for the matrix $G_f$ that is invariant under the group operation of multiplication by lower triangular Toeplitz matrices.

Let $\textsf{G}\subset \mathbb{R}^{i\times i}$ be one such group and consider two matrices $\textsf{A}, \textsf{B}\in \textsf{G}$. Let $\textsf{C} = \textsf{A}\textsf{B}$. Using the fact that $a_{k,l}=a_{k+m,l+m}$ for $\textsf{A} = [a_{k,l}]\in\textsf{G}$, we have that the last row of $\textsf{C}$ is given by
\begin{equation}
c_{il} = \sum_m a_{im}b_{ml} = \sum_m a_{i,m}b_{i,l-m+i}
\enspace.
\end{equation}
If we parametrize these matrices using their lowest row, we can compute the Jacobian for right multiplication as
\begin{equation}
J^R_{[l,m]}=\frac{\partial c_{i,l}}{\partial a_{i,m}} = b_{i,l-m+i}
\enspace.
\end{equation}
Given the triangular structure of $\textsf{B}$, we have that $|J^R|=|b_{i,i}|^i$. As for left multiplication, 
\begin{equation}
J^L_{[l,m]}=\frac{\partial c_{i,l}}{\partial b_{i,m}} = a_{i,l-m+i}
\enspace
\end{equation}
and $|J^L|=|a_{i,i}|^i$. With this we can define the left and right invariant Haar measure
\begin{equation}
\mu(\mathcal{A})=\int_\mathcal{A} \frac{da_{i,\cdot}}{|a_{i,i}|^i}
\enspace.
\end{equation}
Indeed, to check right invariance, let $\textsf{C}=\textsf{A}\textsf{B}$ and observe that
\begin{equation}
\int f(\textsf{C}\textsf{B}^{-1}) ~ \frac{dc_{i,\cdot}}{|c_{i,i}|^i} = \int f(\textsf{A}) ~ \frac{|b_{i,i}|^ida_{i,\cdot}}{|c_{i,i}|^i} = \int f(\textsf{A}) ~ \frac{da_{i,\cdot}}{|a_{i,i}|^i}
\enspace,
\end{equation}
where we used the above defined Jacobian in the first equality and, in the second equality, we used the fact $c_{i,i}=a_{i,i}b_{i,i}$. Left invariance may be checked similarly for $f(\textsf{B}^{-1}\textsf{C})$. Therefore, we have arrived at a prior that is invariant under the group operation.

In the sequence we derive the posterior that corresponds to this prior. Recall that the residues are
\begin{equation}
\mathcal{E} = Y_f-H_{fp}Z_p-H_fU_f = G_f E_f
\enspace.
\end{equation}
Taking the vectorization operation and making use of the noise vector $\bar{e}_f$ on the subspace of dimension $i+j-1$, we have that
\begin{equation}
\text{vec}(\mathcal{E}) = (I\otimes G_f) B_W \bar{e}_f
\enspace.
\end{equation}
Therefore, the residues covariance is given by
\begin{equation}
\Sigma_\mathcal{E} = (I\otimes G_f) B_W B_W^T (I\otimes G_f^T)
\enspace.
\end{equation}
Since $\text{rank}(B_WB_W^T)$ is $i+j-1$, $\Sigma_\mathcal{E}$ is rank deficient and we shall make use of its pseudo-determinant (product of non-zero singular values) in obtaining its pdf:
\begin{equation}
|\Sigma_\mathcal{E}|_+ = |(I\otimes G_f) B_W B_W^T (I\otimes G_f^T)|_+ = |(I\otimes G_f)\cdot \text{chol}(B_W B_W^T)|_+^2
\propto G_{f[i,i]}^{2(i+j-1)}
\enspace,
\end{equation}
where we used the fact that $I\otimes G_f$ and $\text{chol}(B_WB_W^T)$ are lower triangular and the fact that $\text{chol}(B_WB_W^T)$ must have exactly $i+j-1$ nonzero entries on its diagonal. In order to make the vectorization $G_f$ explicit in the likelihood function, we note that
\begin{equation}
E_f = G_f^{-1}\mathcal{E} \Rightarrow B_W \bar{e}_f = (\mathcal{E}^T\otimes I_i) \text{vec}(G_f^{-1}) \Rightarrow 
\bar{e}_f = B_W^\dagger(\mathcal{E}^T\otimes I_i)B_T (G_{f[i,\cdot]}^{-1})^T
\enspace.
\end{equation}
Therefore,
\begin{equation}
p(\mathcal{E}|G_f) \propto \frac{1}{|G_{f[i,i]}|^{i+j-1}}\exp\left(-\frac{1}{2}G_{f[i,\cdot]}^{-1}B_T^T(\mathcal{E}\otimes I_i) (B_W^\dagger)^TB_W^\dagger(\mathcal{E}^T\otimes I_i)B_T (G_{f[i,\cdot]}^{-1})^T\right)
\end{equation}
and the posterior would be proportional to
\begin{align}
p(\mathcal{E}|G_f^{-1})d\mu(G_f^{-1}) &\propto |G_{f[i,i]}^{-1}|^{i+j-1}\exp\left(-\frac{1}{2}G_{f[i,\cdot]}^{-1}\Omega(G_{f[i,\cdot]}^{-1})^T\right)\cdot \frac{1}{|G_{f[i,i]}^{-1}|^i} \\
&= |G_{f[i,i]}^{-1}|^{j-1}\exp\left(-\frac{1}{2}G_{f[i,\cdot]}^{-1}\Omega(G_{f[i,\cdot]}^{-1})^T\right)
\enspace.
\end{align}
Defining the change of varibles $\nu^T=G_{f[i,\cdot]}^{-1}\Omega_L$, we have
\begin{equation}
p(\nu) \propto |\nu_i|^{j-1}\exp\left(-\frac{1}{2}\nu^T\nu\right) = |\nu_i^2|^{\frac{j+1}{2}-1}\exp\left(-\frac{1}{2}\nu^T\nu\right)
\enspace. 
\end{equation}
Therefore $\nu_i^2\sim \chi^2_{j+1}$ and $\nu_l\sim N(0,1)$, for $l=1,\ldots,i-1$. This is precisely the posterior given by equations (\ref{eq:posti})-(\ref{eq:postf}).

If, on the other hand, we assumed that $\E[e_fe_f^T]=I_{ij}$, then
\begin{equation}
|\Sigma_\mathcal{E}|\propto |G_{f[i,i]}|^{2ij}
\end{equation}
and the posterior of equations (\ref{eq:posti2})-(\ref{eq:postf2}) would follow.
\bibliographystyle{unsrt}
\bibliography{RobustBayes}

\end{document}